\documentclass[pdflatex,sn-mathphys-ay]{sn-jnl}

\usepackage{graphicx}
\usepackage{amsmath,amssymb,amsfonts}
\usepackage{amsthm}
\usepackage{xcolor}
\usepackage{booktabs}
\usepackage{enumitem}
\usepackage{cleveref}
\usepackage{physics}

\theoremstyle{thmstyleone}
\newtheorem{theorem}{Theorem}
\newtheorem{proposition}{Proposition}
\theoremstyle{thmstyletwo}
\newtheorem{remark}{Remark}
\theoremstyle{thmstylethree}

\newcommand{\tot}[1]{I(#1)}
\newcommand{\R}{\mathbb{R}}
\newcommand{\E}{\mathbb{E}}
\newcommand{\assref}[1]{\hyperref[assump:#1]{(#1)}}
\newcommand{\D}{\,\mathrm{d}}
\newcommand{\DT}{\mathcal{D}_T}
\newcommand{\UT}{\mathcal{U}_T}
\newcommand{\fkpp}{\text{Fisher--KPP }}
\newcommand{\cmin}{c_{\min}}
\definecolor{steph}{RGB}{144,12,63}

\crefname{section}{Section}{Sections}
\Crefname{section}{Section}{Sections}

\crefname{figure}{Fig.}{Figs.}
\Crefname{figure}{Figure}{Figures}

\crefname{equation}{Eq.}{Eqs.}
\Crefname{equation}{Equation}{Equations}

\crefname{table}{Table}{Tables}
\Crefname{table}{Table}{Tables}

\crefname{theorem}{Theorem}{Theorems}
\crefname{proposition}{Proposition}{Propositions}
\crefname{remark}{Remark}{Remarks}

\begin{document}
\hypersetup{pdftitle={Invasion speed of cell populations with phase-dependent motility and variable cell-cycle times}, pdfauthor={St\'ephanie M. C. Abo and Ruth E. Baker}}

\title[Cell-cycle structure and invasion speed]{Invasion speed of cell populations with phase-dependent motility and variable cell-cycle times}

\author*[1]{\fnm{St\'ephanie M. C.} \sur{Abo}}\email{stephanie.abo@maths.ox.ac.uk}
\author[1]{\fnm{Ruth E.} \sur{Baker}}\email{ruth.baker@maths.ox.ac.uk}
\affil*[1]{\orgdiv{Mathematical Institute}, \orgname{University of Oxford}, \orgaddress{\city{Oxford}, \country{United Kingdom}}}

\abstract{Whether in development, wound healing or tissue engineering, cell populations invade through a combination of cell movement and cell division. Standard models of invasion assume that all cells move and divide at the same constant rates. In experiments, however, a cell's motility can depend on its cell-cycle phase, and cell-cycle times vary between cells. It is not known when and to what extent these two features affect a population’s invasion speed. We introduce an individual-based model in which the diffusion rate, death rate and cell-cycle progression of each cell depend on its cell-cycle phase and on the local cell density, and we derive the corresponding structured partial differential equation. For pulled invasion fronts, we show that, when cell-cycle progression is deterministic, the invasion speed depends only on the cycle-averaged diffusion and net growth rates, so populations that are motile in different phases can invade at the same speed. Among populations with the same mean cell-cycle time and the same averaged rates, variability in cell-cycle times can only increase the invasion speed. This increase is of second order in the coefficient of variation of the cell-cycle time and below 4\% for measured values. The cell-cycle-time distribution, however, predicts the invasion speed only when it is resolved into phases within which diffusion and death rates are constant. Two populations can share the same cell-cycle-time distribution yet allocate that time differently across phases, which changes their invasion speed. We illustrate these results with data for melanoma cell lines and drug-treated epithelial cells.}



% 92C17 Cell movement (chemotaxis, etc.)
% 35C07 Traveling wave solutions
% 82C31 Stochastic methods (Fokker-Planck, Langevin, etc.)
% 35K57 Reaction-diffusion equations

\maketitle

%--------------------------------------------------------------------------------
\section{Introduction}
\label{sec:intro}
%--------------------------------------------------------------------------------

Cell populations invade new space during tissue development, tissue repair and tumour growth \citep{friedl2009}. The invasion process is driven by the interplay between cell movement and proliferation. Experiments often summarise invasion by the invasion speed, the speed at which the front of the population advances. The classical model of invasion, the \fkpp equation, predicts the invasion speed $2\sqrt{Dr}$ for a population with diffusion rate $D$ and net growth rate $r$ \citep{fisher1937,kpp1937}. This prediction assumes that every cell moves and divides at the same constant rates, at all times.

Real cells are more complex. Cells progress through the cell cycle, a sequence of phases (G1, S, G2 and M) that ends in division. The motility of a cell can depend on its phase, but the direction of this dependence varies between cell types. In several cell lines, including the human cervix carcinoma cell line HeLa and the mouse fibroblastoid cell line L929, cells move less in G2 than in G1/S \citep{walmod2004}, and cancer cells are mostly in G1 \citep{yano2014}. In contrast, \citet{kagawa2013cell} found cancer cells in S/G2/M to be more motile and invasive than cells in G1. In three melanoma cell lines, \citet{vittadello2020} reported similar motility in G1 and in S/G2/M. This growing body of research shows that cancer cell motility and invasive properties are closely associated with the cell cycle. Cell-cycle progression also depends on the environment of a cell. In crowded epithelia, a checkpoint at the G1/S transition slows or arrests cell-cycle progression \citep{streichan_2014}. Such phase-dependent behaviour can now be measured in single cells using the fluorescent ubiquitination-based cell-cycle indicator (FUCCI), which labels cells in G1 red and cells in S/G2/M green \citep{sakauesawano2008}. Time-lapse imaging of FUCCI-labelled cells makes it possible to track both phase duration and phase-specific cell motility \citep{haass2014,vittadello2020}.

Cell-cycle times also vary between cells. We call the time from the birth of a cell to its division the cell-cycle time, and its distribution in a population the cell-cycle-time distribution (CCTD). Measured CCTDs are not exponential, as a constant division rate would imply, but peaked around their mean \citep{yates2017,chao2019}. Their spread is commonly quantified by the coefficient of variation (CV) of the cell-cycle time, the ratio of its standard deviation to its mean. Measured CVs typically lie between 0.1 and 0.4 \citep{mosheiff2017}. Yet many models of FUCCI experiments treat each phase as a compartment that cells leave at a constant rate \citep{vittadello2018,simpson2018,simpson2020}, which implies exponentially distributed phase times.

Structured population models relax these assumptions. In a structured model, the population density depends on a structuring variable, such as cell age or cell-cycle phase, in addition to position and time \citep{gurtin_1981,webb1985theory}. These models admit travelling-wave solutions \citep{sowuzou2001,alomarigourley2002}, which describe the front of an advancing population, and have been used to describe tumour invasion and collective cell migration \citep{dyson2007age,lorenzi2025phenotype}. Structured models are usually written at the population level. Their rates are therefore not directly tied to the behaviour of single cells, nor is variability in the rates tied to the measured CV. Deriving the population-level model from a cell-level model can provide these links. Nevertheless, there have been efforts to represent experimental readouts such as CCTDs in structured population-level models. For invasion, \citet{gavagnin2019} showed that the CCTD determines the invasion speed when cell motility does not depend on phase. Still with constant diffusion, \citet{abo2025survival} derived the invasion speed for age-dependent division and death rates and showed that the form of the age-dependent division rate sets the form of the CCTD. Without space, \citet{alshehri2020} showed that the distribution of the times spent in each phase determines the growth rate and the phase composition of a population. It is not known how the invasion speed depends on the cell cycle when motility and death depend on phase, nor how much the measured variability in cell-cycle times matters.

We therefore ask three questions. Does the invasion speed depend on the phase structure of a cell population, e.g., when in the cell cycle cells move or die? Does it depend on the variability of cell-cycle times? Which cell-cycle measurements are needed to predict it? The answers inform experimentalists on what to measure, and modellers on which cell-cycle features a model of invasion must include.

In \cref{sec:model}, we introduce an individual-based model in which the diffusion rate, death rate and cell-cycle progression of each cell depend on its phase and on the local cell density. We derive the corresponding structured partial differential equation formally (Appendix~\ref{app:meanfield}). We then study pulled invasion fronts, whose speed is set by the dynamics at the leading edge, where the cell density is low. In \cref{sec:res1}, we show that, when cell-cycle progression is deterministic, the invasion speed has the \fkpp form with the diffusion and net growth rates averaged over one cycle. The invasion speed is then independent of the phases in which cells move, whereas the phase profile at the leading edge depends on that phase structure. In \cref{sec:res2}, we derive the invasion speed for stochastic cell-cycle progression. We show that variability in cell-cycle times can only increase the invasion speed, and that the increase is of second order in the CV. In \cref{sec:res3}, we show that the CCTD predicts the invasion speed only when it is resolved into phases, and that the effect of a drug that lengthens the cell cycle depends on the phase it lengthens. We illustrate the results with data for FUCCI-labelled melanoma cells \citep{vittadello2020} and drug-treated epithelial cells \citep{jones_2018}, and discuss their implications in \cref{sec:discussion}.

%--------------------------------------------------------------------------------
\section{From cell- to population-level models of collective migration}
\label{sec:model}
%--------------------------------------------------------------------------------
We first define a stochastic individual-based model of interacting cells, then derive a continuum description through a formal mean-field limit, which allows us to study the population-level behaviour of the system.

We describe each cell by its position and its cell-cycle phase. At time $t\geq0$, the population consists of $M(t)$ cells. Cell $i$ has position $X_i(t)\in\R$ and cell-cycle phase $A_i(t)\in[0,1)$, the fraction of the cell cycle completed. A cell is born with phase $A_i=0$ and divides when its phase first reaches $A_i=1$. It is then replaced by two daughter cells at the same position, both with phase $0$. A cell may also die, in which case it is removed. At $t=0$, the $M(0)$ cells have given positions $X_i(0)\in\R$ and phases $A_i(0)\in[0,1)$. We use $x$ and $a$ to denote an arbitrary position and phase, respectively.

We measure position in units of a reference length $L$, such as the width of the experimental domain. Cells have size $\ell\ll L$, such that $K:=L/\ell$ is the number of cells that fit side by side along $L$. Cells interact through the local cell density
\begin{equation}\label{eq:local-density}
    N_\varepsilon(x,t):=\frac{1}{K}\sum_{j=1}^{M(t)}B_\varepsilon(x-X_j(t)),\qquad B_\varepsilon(x):=\frac{1}{\varepsilon}B\left(\frac{x}{\varepsilon}\right),
\end{equation}
where $B$ is a nonnegative, even and smooth function with support on $[-1,1]$, with $\int_\R B(x)\D x=1$. The parameter $\varepsilon>0$ is the radius over which a cell senses its neighbours, relative to $L$. Thus, $N_\varepsilon(x,t)$ is the rescaled cell density averaged over that local neighbourhood.
So, $N_\varepsilon(x,t)=1$ corresponds to a confluent neighbourhood where cells are packed side by side, and $N_\varepsilon(x,t)=0$ to an empty one.

Each cell $i$ diffuses in space and progresses through the cell cycle according to
\begin{align}
    \D X_i &= \partial_x  D(A_i,N_\varepsilon)\,\D t+\sqrt{2D(A_i,N_\varepsilon)}\,\D W_i^x,\label{eq:micro-sde-x} \\
    \D A_i &= v(A_i,N_\varepsilon)\,\D t+\sigma(A_i)\,\D W_i^a,\label{eq:micro-sde-a}
\end{align}
where $N_\varepsilon=N_\varepsilon(X_i(t),t)$, and $W_i^x$ and $W_i^a$ are independent standard Wiener processes in position and cell-cycle phase, respectively, and also independent across cells. The diffusion rate $D(a,N_\varepsilon)$, cycling speed $v(a,N_\varepsilon)$ and death rate $\mu(a,N_\varepsilon)$ depend on the cell-cycle phase and local density, while $\sigma(a)$ depends only on the cell-cycle phase. The latter describes the variability in cell-cycle progression, where each cell follows its own phase path $A(t)$ independently of its mother and its sister. Cell-cycle times therefore vary between cells and between generations. The drift $\partial_x  D$ in \cref{eq:micro-sde-x} is the It\^o correction that gives Fickian diffusion at the population level.

The cell-cycle time of a newborn cell is the first-passage time
\begin{equation}\label{eq:cycle-time}
    T:=\inf\{t>0:A(t)=1\},\qquad A(0)=0.
\end{equation}
At a fixed local density, newborn cells have independent and identically distributed cell-cycle times with probability density $f_T$, which is the CCTD at that density. In what follows, $f_T$ denotes the CCTD at low density, $N=0$, which is the one that sets the invasion speed. When the density changes in time and space, cell-cycle times need not be independent or identically distributed.\\

\noindent\textbf{Assumptions.} The functions $D$, $v$, $\sigma$ and $\mu$ are smooth. For all $a\in[0,1]$ and $N\geq0$,
\begin{itemize}
    \item[\textbf{A1.}]\phantomsection\label{assump:A1} $v(a,N)\geq v_{\min}>0$;
    \item[\textbf{A2.}]\phantomsection\label{assump:A2} $\sigma(a)>0$ for $a\in(0,1)$, with $\sigma(0)=\sigma(1)=0$;
    \item[\textbf{A3.}]\phantomsection\label{assump:A3} $\sigma^2(a)=o(a)$ as $a\to0$;
    \item[\textbf{A4.}]\phantomsection\label{assump:A4} $D(a,N)\geq0$;
    \item[\textbf{A5.}]\phantomsection\label{assump:A5} $\mu(a,N)\geq0$.
\end{itemize}
Assumption A1 means that crowding can slow cell-cycle progression but cannot arrest it. Without death, a cell divides in finite time, with mean at most $1/v_{\min}$. Assumptions A2 and A3 make $a=0$ an entrance boundary, meaning that cells born at $a=0$ start to progress immediately, and cells with $a>0$ do not return to $a=0$.

When cells are small relative to $L$, $K\to\infty$, and interactions are local, $\varepsilon\to0$, the population is described by the density $p(a,x,t)$ of cells with phase $a$ at position $x$ and time $t$. In Appendix~\ref{app:meanfield}, we show formally that $p$ satisfies
\begin{align}
    \partial_t p+\partial_a  J &=\partial_x \left(D(a,N)\partial_x  p\right)-\mu(a,N)p, \label{eq:macro-pde}\\
    J(0,x,t) &=2J(1,x,t), \label{eq:renewal-condition}\\
    p(a,x,0) &= p_0(a,x) \geq 0, \label{eq:initial-condition}
\end{align}
where
\begin{align}
    J(a,x,t) &:=v(a,N)p-\partial_a \left(\Sigma(a)p\right),\qquad \Sigma(a):=\frac{1}{2}\sigma^2(a), \label{eq:phase-flux}\\
    N(x,t) &:=\int_0^1p(a,x,t)\,\D a. \label{eq:local-density-limit} 
\end{align}
The phase flux $J$ is the number of cells crossing phase $a$ per unit time. The renewal condition~\eqref{eq:renewal-condition} accounts for the replacement of each dividing cell by two daughter cells with phase $0$.

We further assume that, for all $a\in[0,1]$ and $N\geq0$,
\begin{equation}\label{eq:density-assumptions}
    v(a,N)\leq v(a,0),\qquad D(a,N)\leq D(a,0),\qquad \mu(a,N)\geq\mu(a,0).
\end{equation}
Crowding slows cell-cycle progression, reduces motility and increases cell death. This density dependence keeps the population bounded behind the invasion front. We consider pulled invasion fronts, whose asymptotic speed is set by the dynamics at the leading edge, where $N\to0$. We denote the corresponding low-density rates by $v_0(a):=v(a,0)$, $D_0(a):=D(a,0)$ and $\mu_0(a):=\mu(a,0)$.

%--------------------------------------------------------------------------------
\section{Cycle-averaged rates determine the invasion speed}\label{sec:res1}
%--------------------------------------------------------------------------------

We first consider deterministic cell-cycle progression with $\Sigma \equiv 0$. The phase flux \eqref{eq:phase-flux} then reduces to $J=v(a,N)p$. In this case, cells at a fixed density take the same amount of time to complete a cycle. 

Far behind the wave front, the population reaches a homogeneous equilibrium $\bar p(a)$, which satisfies
\begin{equation}\label{eq:pbar}
    \bar p(a)=\frac{v(0,\bar N)\bar p(0)}{v(a,\bar N)}\exp\left(-\int_0^a\frac{\mu(s,\bar N)}{v(s,\bar N)}\D s\right), \qquad \bar N=\int_0^1\bar p(a)\D a.
\end{equation}
The renewal condition then requires $e^{-\Phi(\bar N)}=1/2$, where
\begin{equation}\label{eq:phi}
    \Phi(N):=\int_0^1\frac{\mu(a,N)}{v(a,N)}\D a
\end{equation}
is the cumulative death hazard over one cell cycle at a given density $N$. That is, exactly half of the cells must survive to division. Since $\Phi$ is continuous, an equilibrium exists whenever $\Phi(0)<\ln2$ and $\Phi(N)>\ln 2$ for some $N>0$. This equilibrium is unique when $\Phi$ is strictly increasing.

At the leading edge, cells spend a time $\D a/v_0(a)$ in a small phase interval $\D a$, so the time integral of a rate $h(a)$ over one cell cycle is
\begin{equation}\label{eq:cycle-total}
    \tot{h}:=\int_0^1\frac{h(a)}{v_0(a)}\D a.
\end{equation}
In particular, $\tau:=\tot{1}$ is the cell-cycle duration, $2\tot{D_0}$ is the mean-square displacement of a cell over one cycle and $\tot{\mu_0}$ is its cumulative death hazard, so that $e^{-\tot{\mu_0}}$ is the probability that a cell survives to division. We write $\bar h:=\tot{h}/\tau$ for the time average of $h$ over the cell cycle. For example, $\bar D_0$ is the averaged diffusion rate.

In the travelling-wave coordinate $z=x-ct$, $c>0$, the model reduces to the linear problem
\begin{equation}\label{eq:edge-problem}
\begin{aligned}
    -c\,\partial_z p+\partial_a \left(v_0(a)p\right)&=D_0(a)\partial_{zz}p-\mu_0(a)p,\\
    v_0(0)p(0,z)&=2v_0(1)p(1,z).
\end{aligned}
\end{equation}
We look for solutions of the form $p(a,z)=\varphi(a)e^{-\lambda z}$, where $\lambda>0$ is the spatial decay rate of the leading edge and $\varphi(a)$ is its phase profile. Substituting into \cref{eq:edge-problem}, we obtain
\begin{equation}\label{eq:edge-ode}
    \partial_a \left(v_0(a)\varphi\right)=\left(D_0(a)\lambda^2-c\lambda-\mu_0(a)\right)\varphi.
\end{equation}
Integrating over one cycle and applying the renewal condition gives the dispersion relation
\begin{equation}\label{eq:F0}
    F_0(c,\lambda):=\tot{D_0}\lambda^2-c\tau\lambda-\tot{\mu_0}+\ln 2=0.
\end{equation}
\begin{theorem}\label{thm:speed}
Suppose $\tot{D_0}>0$ and $\tot{\mu_0}<\ln2$. Then \cref{eq:edge-problem} has a positive solution of the form $\varphi(a)e^{-\lambda z}$ with $\lambda>0$ if and only if $c\geq\cmin$, where
\begin{equation}\label{eq:cmin}
    \cmin=2\sqrt{\bar D_0 \bar r_0},\quad \text{with}\quad \bar D_0=\frac{\tot{D_0}}{\tau},\qquad \bar r_0=\frac{\ln2-\tot{\mu_0}}{\tau}.
\end{equation}
The minimum is attained at $\lambda_c=\sqrt{\bar r_0/\bar D_0}$.
\end{theorem}

The theorem follows from the discriminant of \cref{eq:F0} (Appendix~\ref{app:speed-deterministic}). The invasion speed has the \fkpp form, where $\bar D_0$ and $\bar r_0$ are the low-density cycle-averaged diffusion and net growth rates, respectively, so phase structure affects the invasion speed only through these averaged rates. Cell populations with different phase structures therefore invade at the same speed whenever the product $\bar D_0\bar r_0$ is the same. The same holds for a single population before and after a change in its phase structure that preserves this product. The condition $\tot{\mu_0}<\ln2$ means that more than half of the cells survive to division (\cref{eq:phi}); since $\tot{\mu_0}=\Phi(0)$, it is also the condition for the population to persist.

\begin{figure}[ht]
\centering
\includegraphics[width=\linewidth]{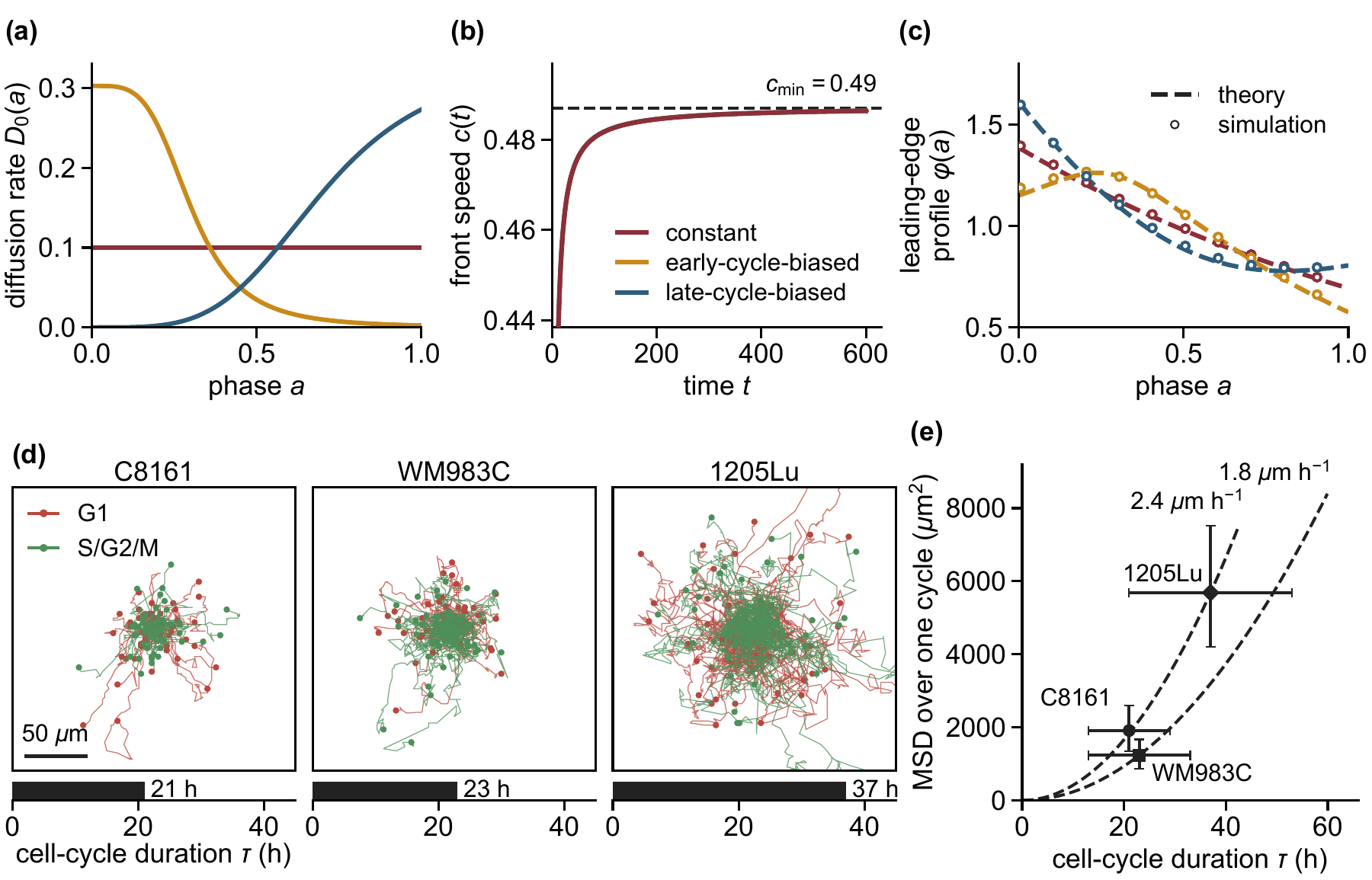}
\caption{Cell populations with the same averaged rates $\bar D_0$ and $\bar r_0$ invade at the same speed. (a) Phase-dependent diffusion profiles for constant, early-cycle-biased and late-cycle-biased motility, with $v_0(a)\equiv 1$ and $\mu_0(a)\equiv 0.1$ in all three cases. (b) Invasion speeds from cases in (a) converge to $\cmin$ (dashed line). (c) Normalised phase profile of the leading edge from \cref{eq:edge-profile} (dashed lines) and from numerical simulations (circles). (d) Trajectories of 50 cycling cells in G1 (red) and 50 in S/G2/M (green) for three melanoma cell lines. Trajectories are translated to start at the origin. Bars show the cell-cycle duration $\tau$. (e) Mean-square displacement along one axis over one cycle, $2\tot{D_0}$, as a function of $\tau$. Dashed curves show combinations with equal $\cmin$. Data from \citet{vittadello2020}; error bars as described in Appendix~\ref{app:data-analysis}.}
\label{fig:averages}
\end{figure}

\Cref{fig:averages} illustrates this result. We simulated three cell populations with different diffusion rates $D_0(a)$ but the same $\tau$, $\tot{D_0}$ and $\tot{\mu_0}$ (\cref{fig:averages}a). The three invasion fronts converge to the same speed $\cmin$ (\cref{fig:averages}b), but their phase profiles at the leading edge differ (\cref{fig:averages}c). This is because the invasion speed depends only on time integrals over the whole cycle, whereas leading-edge phase profiles depend on how the rates vary within the cycle.

Solving \cref{eq:edge-ode} with $c=\cmin$ and $\lambda=\lambda_c$ we obtain
\begin{equation}\label{eq:edge-profile}
\varphi(a)\propto \frac{1}{v_0(a)}\exp\!\left(\int_0^a \frac{(\bar r_0/\bar D_0)D_0(s)-2\bar r_0-\mu_0(s)}{v_0(s)}\D s\right).
\end{equation}
This profile depends on the phase structure of the diffusion rate $D_0(a)$ through its cumulative integral $\int_0^a D_0(s)/v_0(s)\D s$ as well as on that of the death rate $\mu_0(a)$. In addition, the factor $1/v_0(a)$ reflects the time that cells spend in each phase. Hence, the phase profile is weighted towards phases in which cells progress slowly. The phase profile can therefore provide information on how the rates change across the cell cycle. 

We compare $\varphi(a)$ with the profile $\varphi_{\rm const}(a)$ of a population with constant diffusion $\bar D_0$ and the same $v_0(a)$ and $\mu_0(a)$. We obtain
\begin{equation}\label{eq:edge-ratio}
\frac{\D}{\D a}\ln\frac{\varphi(a)}{\varphi_{\rm const}(a)}=\frac{\bar r_0}{\bar D_0}\,\frac{D_0(a)-\bar D_0}{v_0(a)}.
\end{equation}
Relative to a phase profile with a constant diffusion rate, $\varphi$ increases where $D_0(a)$ is above its average and decreases where it is below. In \cref{fig:averages}c, the profile for early-cycle-biased motility rises relative to the constant case for early phases and decreases afterwards, whereas the profile for late-cycle-biased motility falls until $a\approx0.6$ and increases afterwards.

\Cref{fig:averages}d,e illustrate, with data for three FUCCI-labelled melanoma cell lines \citep{vittadello2020}, the limitation of measuring only the invasion speed to differentiate between cell populations. In these experiments, cells move freely at low density, so the measured rates would reflect those that set the invasion speed at the leading edge. C8161 and WM983C cells have similar cell-cycle durations, but WM983C cells move less, so their predicted invasion speed is lower, although the error bars in \cref{fig:averages}e overlap and the data do not resolve this difference. The 1205Lu cells move more than C8161 cells but have a longer cell cycle. Because these two effects compensate, C8161 and 1205Lu cells have nearly the same predicted invasion speed, 2.45 and $2.40\,\mu$m\,h$^{-1}$, respectively. The invasion speed alone cannot tell these two cell lines apart, even though their diffusion and proliferation rates differ.

%--------------------------------------------------------------------------------
\section{The invasion speed is robust to cell-cycle-time variability}
\label{sec:res2}
%--------------------------------------------------------------------------------

Real cell-cycle times vary, and so we need to consider the impact of this variability on the invasion speed. We next investigate whether cycle-averaged rates still determine the invasion speed when cell-cycle progression is stochastic, that is, $\Sigma>0$.

Each cell follows its own phase path $A(t)$, given by \cref{eq:micro-sde-a}, independently of its mother and its sister. At low density, the cell-cycle time $T$ of a newborn cell is the first-passage time in \cref{eq:cycle-time}, with cycling speed $v_0(a)$. At the leading edge, the phase flux is now $J=v_0(a)p-\partial_a \left(\Sigma(a)p\right)$, and the linear problem becomes
\begin{equation}
    \label{eq:edge-problem-sigma}
\begin{aligned}
    -c\,\partial_z p+\partial_a  J&=D_0(a)\partial_{zz}p-\mu_0(a)p, \qquad z=x-ct, \quad c>0,\\
    J(0,z)&=2J(1,z),\\
  J(a,z)&=v_0(a)p-\partial_a \left(\Sigma(a)p\right).
\end{aligned}
\end{equation}
We again look for solutions of the form $p(a,z)=\varphi(a)e^{-\lambda z}$. The phase profile now solves a second-order equation in $a$, which has no closed-form solution in general. The renewal condition, however, can still be written in terms of the phase paths of single cells, using the Feynman--Kac formula (Appendix~\ref{app:speed-stochastic}).
\begin{proposition}
    \label{prop:stoch-speed}
Suppose $\tot{D_0}>0$ and $2\,\E\big[\exp\big(-\int_0^T\mu_0(A(t))\D t\big)\big]>1$. If \cref{eq:edge-problem-sigma} has a positive solution of the form $\varphi(a)e^{-\lambda z}$ with $\lambda>0$, then $c=c(\lambda)$, where
\begin{equation}\label{eq:disp-stoch}
    2\,\E\left[\exp\left(\lambda^2\int_0^T D_0(A(t))\D t-\int_0^T\mu_0(A(t))\D t-c(\lambda)\lambda T\right)\right]=1.
\end{equation}
The expectation is over the phase path $A(t)$ of a newborn cell at low density, with $A(0)=0$. The invasion speed is $c_\ast:=\min_{\lambda>0}c(\lambda)$.
\end{proposition}

As in \cref{sec:res1}, the quantity $2\int_0^T D_0(A(t))\D t$ is the mean-square displacement of a newborn cell between birth and division, given its phase path, and $\int_0^T\mu_0(A(t))\D t$ is the cumulative death hazard over its cycle. The condition on $\mu_0$ means that a newborn cell survives to division with probability greater than 1/2. This condition is the analogue of the persistence condition $\tot{\mu_0}<\ln2$ in \Cref{thm:speed}. As $\Sigma\to0$, all cells take the same time $T=\tau$, and the two integrals tend to $\tot{D_0}$ and $\tot{\mu_0}$. Equation~\eqref{eq:disp-stoch} then reduces to $F_0=0$ in \cref{eq:F0}, and $c_\ast=\cmin$.

\Cref{prop:stoch-speed} gives the exact invasion speed, but $c_\ast$ must be computed numerically. In practice, this $c_\ast$ depends on how cell-cycle times, displacements and survival vary together between cells, which is rarely measured. Experiments usually report this variability through the CV of the cell-cycle time, measured by lineage tracking. To relate $c_\ast$ to the measured CV, we expand \cref{eq:disp-stoch} for small phase noise about the solution $(\cmin,\lambda_c)$ of \Cref{thm:speed}. We compare populations with the same mean cell-cycle time and the same averaged diffusion and death rates (Appendix~\ref{app:cv}). This is the relevant regime, since measured CVs typically lie between 0.1 and 0.4 \citep{mosheiff2017,cadart2018}. 
\begin{proposition}\label{prop:cv}
Suppose $\Sigma(0)=\Sigma(1)=0$, as in Assumption A2. Let $\tau$, $\tot{D_0}$ and $\tot{\mu_0}$ in \cref{eq:cmin} be replaced by their means over one cycle, $\E[T]$, $\E\big[\int_0^T D_0(A(t))\D t\big]$ and $\E\big[\int_0^T \mu_0(A(t))\D t\big]$, and let $\cmin$ and $\lambda_c$ be given by \cref{eq:cmin}. Then, for small $\Sigma$,
\begin{equation}\label{eq:cv-general}
    c_\ast = \cmin + \frac{F_1(\cmin,\lambda_c)}{\tau\lambda_c}+ O(\Sigma^2), \qquad F_1(c,\lambda):=\int_0^1\frac{\Sigma(a)}{v_0(a)}\,g_0(a)^2\D a,
\end{equation}
where $g_0(a) := \left(D_0(a)\lambda^2 - c\lambda-\mu_0(a)\right)/v_0(a)$.
\end{proposition}
The correction $F_1$ is of order $\Sigma$, as is $\mathrm{CV}^2$ (Appendix~\ref{app:cv}), so the correction to the invasion speed is of second order in the CV. Since $F_1\geq0$, variability in cell-cycle times can only increase the invasion speed. This remains true when the phase noise is not small: provided that the mean cumulative death hazard over one cycle is below $\ln2$, Jensen's inequality gives $c_\ast\geq\cmin$ for any $\Sigma$ (Appendix~\ref{app:cv}). Among populations with the same mean cell-cycle time and the same averaged diffusion and death rates, the population with deterministic cell-cycle progression therefore invades the slowest. The averaged rates of \Cref{thm:speed} still set the invasion speed when cell-cycle times vary, up to a small increase, and within the range of measured CVs the invasion speed should barely depend on cell-cycle-time variability.

For example, with constant rates $v_0(a)\equiv1/\tau$, $D_0(a)\equiv D$ and $\mu_0(a)\equiv\mu$, \cref{eq:cv-general} becomes
\begin{equation}
    \label{eq:cv-constant}
    \frac{c_\ast}{\cmin}= 1 + \frac{(\ln2)^2}{4\left(\ln2 - \tau\mu\right)}\,\mathrm{CV}^2+O(\mathrm{CV}^4).
\end{equation}
When the death rate is low ($\mu\approx 0$), as is often the case in low-density experiments, and for typical CVs below 0.4, the correction to the invasion speed is less than 3\%. 

\begin{figure}[h]
\centering
\includegraphics[width=\linewidth]{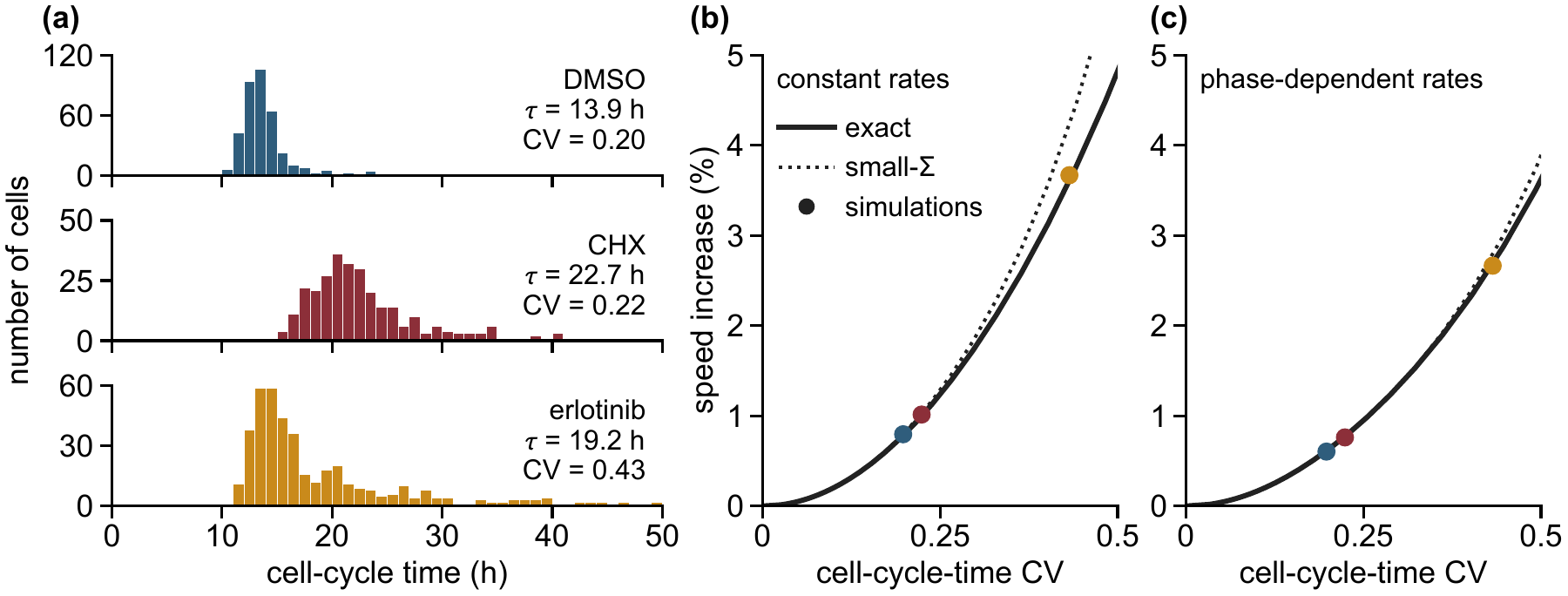}
\caption{Invasion speed is robust to cell-cycle-time variability. (a) Measured CCTDs for DMSO, CHX and erlotinib \citep{jones_2018}. (b, c) Relative increase in invasion speed, $c_\ast/\cmin - 1$, as a function of the CV, for (b) constant and (c) phase-dependent rates, from \Cref{prop:stoch-speed} (solid), \Cref{prop:cv} (dotted) and numerical simulations (circles, coloured as in (a)). In (b) and (c), the mean cell-cycle time and the averaged diffusion and death rates do not change with the CV.}
\label{fig:variability}
\end{figure}

We now test these predictions for experimentally measured CCTDs. We consider lineage-tracking data for control (DMSO) and drug-treated cells \citep{jones_2018}. Cycloheximide (CHX) inhibits protein synthesis and erlotinib inhibits the epidermal growth factor receptor (EGFR). Both drugs lengthen the cell cycle. The measured CVs are 0.20 for DMSO, 0.22 for CHX and 0.43 for erlotinib (\cref{fig:variability}a). For each condition, we simulate our model (\crefrange{eq:macro-pde}{eq:initial-condition}) with $\Sigma(a)=\Sigma_0a^2(1-a)^2$, and choose $\Sigma_0$ to match the measured CV. We compare the invasion speed obtained numerically with our theoretical predictions for illustrative constant and phase-dependent rates (\cref{fig:variability}b,c).

For constant rates, cell-cycle-time variability increases the invasion speed by 0.8\% for DMSO, 1.0\% for CHX and 3.7\% for erlotinib (\cref{fig:variability}b). With phase-dependent rates, the increase is 0.6\%, 0.8\% and 2.7\%, respectively (\cref{fig:variability}c). The averaged rates predict the invasion speed to within 4\%, whether or not the rates depend on phase. For the three melanoma cell lines of \cref{fig:averages}, for which \citet{vittadello2020} report the CV of the cell-cycle time, the predicted invasion speeds increase by at most 3.3\% (Appendix~\ref{app:data-analysis}).

%--------------------------------------------------------------------------------
\section{Predicting the invasion speed from cell-cycle measurements}\label{sec:res3}
%--------------------------------------------------------------------------------
\Cref{sec:res1,sec:res2} show that the invasion speed depends on the cell cycle mainly through averaged rates. In this section, we ask which measurements of the cell cycle are needed to predict the invasion speed. The typical measurement is the CCTD, which we focus on here. We first show that the CCTD is enough when the diffusion and death rates are constant, but not when they depend on phase, unless it is resolved by phase. We then illustrate the difference with a treatment that lengthens one phase of the cell cycle.

\subsection{The CCTD predicts the invasion speed when resolved by phase}\label{sec:res3-cctd}

Models of cell invasion have related the CCTD to the invasion speed when the diffusion rate of cells does not depend on cell-cycle phase \citep{abo2025survival}; \citet{gavagnin2019} also assumed a constant death rate. Given phase-dependent diffusion and death rates, is the CCTD sufficient to predict the invasion speed?

We consider the dispersion relation \eqref{eq:disp-stoch} of \Cref{prop:stoch-speed}. When the diffusion and death rates are constant, $D_0(a)\equiv D$ and $\mu_0(a)\equiv \mu$, the phase path enters only through the cell-cycle time $T$, and the dispersion relation reduces to
\begin{equation}\label{eq:disp-cctd}
    2\,\mathcal L\{f_T\}(s)=1,\qquad s:=c\lambda-D\lambda^2 + \mu,
\end{equation}
where $\mathcal L\{f_T\}(s):=\E[e^{-sT}]$ is the Laplace transform of the CCTD $f_T$ (Appendix~\ref{app:speed-stochastic}). The invasion speed is then $c_\ast=2\sqrt{D(s^\ast - \mu)}$, where $s^\ast$ is the unique solution of \cref{eq:disp-cctd}. So, when the rates are constant, the CCTD is the only information about the cell cycle needed to predict the invasion speed. If the diffusion and death rates of two populations are known, their CCTDs are enough to determine which one invades faster. This is the result of \citet{gavagnin2019}. For example, deterministic progression (CV = 0) gives $s^\ast = \ln2/\tau$, and an exponential CCTD (CV = 1) gives $s^\ast = 1/\tau$. The more variable population invades faster, by a factor $1/\sqrt{\ln 2}$, as follows from the results of \citet{gavagnin2019}.

When the diffusion rate or the death rate varies through the cell cycle, the CCTD is no longer enough. The invasion speed depends on the joint distribution of $T$, $\DT:=\int_0^T D_0(A(t))\D t$ and $\UT:=\int_0^T \mu_0(A(t))\D t$, that is, on how long a cell takes to divide, how much it diffuses while doing so, and how likely it is to survive until division. Both $\DT$ and $\UT$ depend on how long a cell spends in each phase, whereas the CCTD records only $T$. For illustration, consider two populations with the same diffusion rate in G1, and a different but shared diffusion rate in S/G2/M. In the first population, G1 is long and S/G2/M is short. In the second, the times spent in the two phases are exchanged. Both populations have the same CCTD, because the total time $T$ is the same. But if cells are more motile in G1, the first population spends more time in a phase where they move fast, and invades faster. 

Suppose that the cell cycle consists of phases $i=1,\dots,n$, such as G1 and S/G2/M resolved by the two-colour FUCCI reporter. Each phase is an interval of the phase variable $a$. Suppose also that $D_i$ and $\mu_i$ are constant within phase $i$. If $T_i$ is the total time a cell spends in phase $i$, then the mean-square displacement over one cycle, given the phase times, is $2\sum_i D_i T_i$ and the cumulative death hazard is $\sum_i \mu_i T_i$. The dispersion relation \eqref{eq:disp-stoch} becomes
\begin{equation}
    \label{eq:disp-phases}
    2\,\mathcal L\{f_{T_1,\dots,T_n}\}(s_1,\dots,s_n)=1,\qquad s_i:=c\lambda-D_i\lambda^2+\mu_i,
\end{equation}
where $\mathcal L\{f_{T_1,\dots,T_n}\}(s_1,\dots,s_n):=\E\big[e^{-s_1T_1-\cdots-s_nT_n}\big]$ is the Laplace transform of the joint distribution of the phase times. By a CCTD resolved by phase, we mean this joint distribution. When the phase times are independent, \cref{eq:disp-phases} becomes
\begin{equation}
    \label{eq:disp-phases-indep}
    2\prod_{i=1}^n\mathcal L\{f_{T_i}\}(s_i)=1,
\end{equation}
where $f_{T_i}$ is the distribution of times spent in phase $i$. Each phase enters with its own $s_i$, i.e., with its own diffusion and death rates. \Cref{eq:disp-phases} reduces to \cref{eq:disp-cctd} only when $s_i=s$ for every phase, which occurs when the rates are constant. The CCTD remains an important experimental readout and is informative at the resolution of the cell cycle at which the rates are constant.

In practice, the mean phase durations matter most. If we keep only the means in \cref{eq:disp-phases}, we obtain $\sum_i s_i\,\E[T_i]=\ln2$, which is the dispersion relation \eqref{eq:F0} of \Cref{thm:speed}. The variability of the phase times changes the invasion speed only slightly, as in \cref{sec:res2}. So, to predict the invasion speed, we need the mean time spent in each phase and the diffusion and death rates in each phase. Combining time-lapse imaging of FUCCI-labelled cells with single-cell tracking provides both.

There are two considerations to keep in mind. First, the times spent in successive phases can be correlated. The joint distribution in \cref{eq:disp-phases} is then needed. Second, \cref{eq:disp-phases} assumes that the diffusion and death rates are constant within each phase. If they vary within S/G2/M, for example, a finer resolution of the cycle is needed. The FUCCI4 reporter, which resolves G1, S, G2 and M, could be used to achieve this \citep{bajar2016fucci4}.

\subsection{A practical example: lengthening a cell-cycle phase}\label{sec:res3-lengthening}

We illustrate a practical consequence for drugs that lengthen the cell cycle. The CCTDs of treated cells are commonly recorded, such as those of CHX and erlotinib in \cref{sec:res2} \citep{jones_2018, leander2014}. These CCTDs give the increase in mean cell-cycle time, $\Delta t$, but they usually come from lineage studies that do not measure motility, so the diffusion rate may be known only for untreated cells. If treated cells were tracked, their averaged diffusion rate would be known, and \Cref{thm:speed} would give their invasion speed directly. We therefore ask how well the invasion speed of treated cells can be predicted from $\Delta t$ and from measurements on untreated cells. Without phase resolution, single-cell tracking of untreated cells gives only the cycle-averaged diffusion rate $\bar D_0$. With phase resolution, FUCCI imaging combined with single-cell tracking also gives the mean duration and the diffusion rate of each phase.

Let $\tau_i:=\E[T_i]$ be the mean duration of phase $i$, so that $\tau=\sum_i\tau_i$. Suppose that a drug lengthens phase $k$ by an additional time $\Delta t$ on average, without changing the diffusion and death rates in each phase. Then $\tau$, $\tot{D_0}=\sum_i D_i \tau_i$ and $\tot{\mu_0}=\sum_i \mu_i \tau_i$ increase by $\Delta t$, $D_k\Delta t$ and $\mu_k\Delta t$, respectively. Substituting into \cref{eq:cmin} gives
\begin{equation}
\label{eq:lengthening}
\frac{\cmin(\Delta t)}{\cmin(0)}=\frac{\tau}{\tau+\Delta t}\sqrt{\left(1+\frac{\Delta t}{\tau}\frac{D_k}{\bar D_0}\right)\left(1-\frac{\Delta t}{\tau}\frac{\mu_k}{\bar r_0}\right)}.
\end{equation}
\Cref{eq:lengthening} is exact for deterministic cell-cycle progression. With variable cell-cycle times, it holds up to the small correction of \cref{sec:res2}. The change in invasion speed depends on the increase $\Delta t$, which the CCTD of treated cells gives, on the phase $k$ that is lengthened, and on the diffusion and death rates in that phase relative to their cycle averages. The phase that is lengthened is often known from the mechanism of the drug. For example, a drug that delays the G1/S transition lengthens G1. The rates in that phase, however, require phase-resolved imaging.

Without such phase resolution, the only diffusion rate available is $\bar D_0$, so we set $D_k=\bar D_0$ in \cref{eq:lengthening}. When death is negligible, $\mu_k\ll\bar r_0$, this gives
\begin{equation}
\label{eq:lengthening-cctd}
\frac{\cmin(\Delta t)}{\cmin(0)}=\sqrt{\frac{\tau}{\tau+\Delta t}},
\end{equation}
whichever phase is lengthened. This prediction is correct when cells in the lengthened phase have the average diffusion rate, $D_k=\bar D_0$. It also holds when all phases lengthen in proportion, since then $\bar D_0$ is unchanged and only $\tau$ increases. Otherwise, the prediction overestimates the reduction in invasion speed when cells in that phase are more motile than average, and underestimates it when they are less motile than average.

\begin{figure}[ht]
\centering
\includegraphics[width=\textwidth]{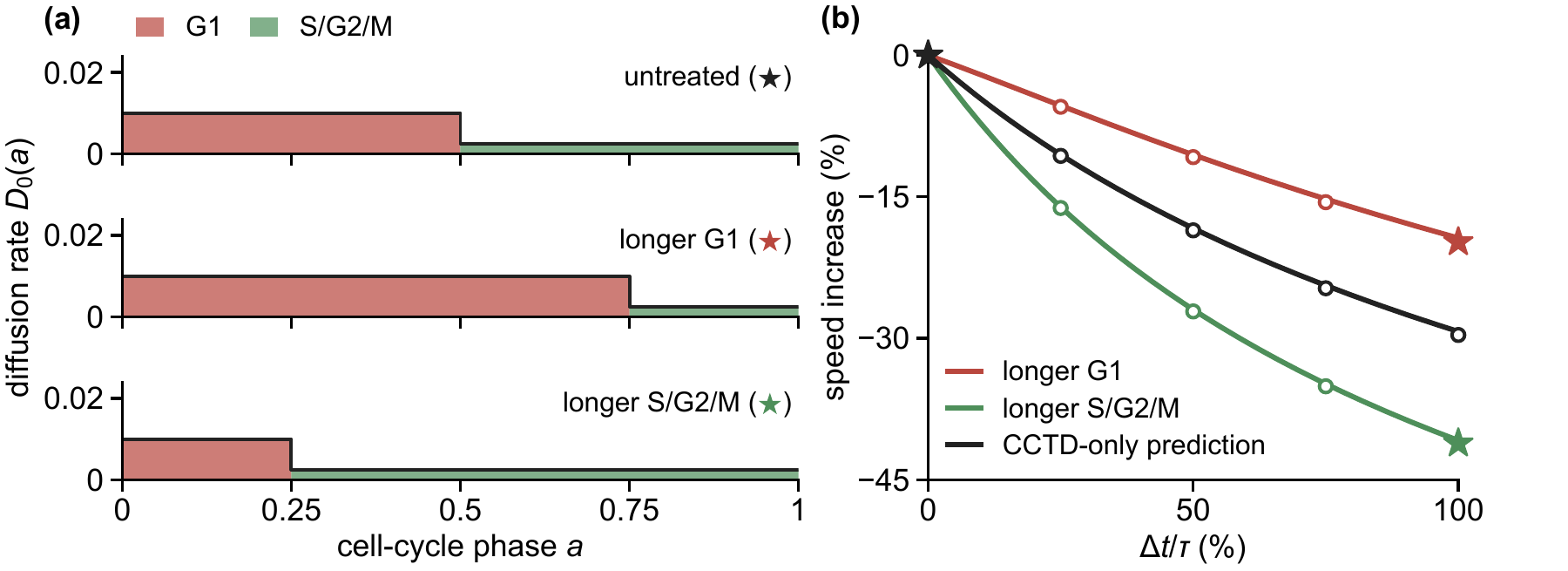}
\caption{Effects of lengthening the duration of a cell-cycle phase. (a) Diffusion rate $D_0(a)$ before treatment (black star) and after doubling the cell-cycle duration by lengthening G1 (red star) or S/G2/M (green star). (b) Percentage change in invasion speed as a function of the relative increase in cell-cycle duration, $\Delta t/\tau$, from \cref{eq:lengthening} (lines) and numerical simulations (circles). Stars mark the cases in (a). The black line is the prediction from the CCTD alone, \cref{eq:lengthening-cctd}, which coincides with lengthening all phases in proportion. Before treatment, $\tau=1$ and the low-density rates are $v_0(a)=1$, $D_0(a)=0.01$ for $a<1/2$ (G1) and $D_0(a)=0.0025$ for $a\geq1/2$ (S/G2/M), and $\mu_0(a)=0$.}
\label{fig:lengthening}
\end{figure}

In \cref{fig:lengthening}, G1 lasts for half of the cell-cycle duration, and G1 cells are four times as motile as S/G2/M cells. For a drug that doubles the cell-cycle duration, the CCTD alone predicts a 29\% reduction in invasion speed. The reduction is only 19\% when the drug lengthens G1, but 41\% when it lengthens S/G2/M. So the same change in the CCTD can correspond to very different changes in invasion speed, depending on the phase that is lengthened.

Without phase resolution, \cref{eq:lengthening-cctd} always predicts that a drug that lengthens the cell cycle slows invasion. With phase resolution, \cref{eq:lengthening} shows that this is not necessarily the case. Lengthening any phase lowers the net growth rate $\bar r_0$, because cells divide less often and are susceptible to death for longer. However, lengthening a phase in which cells are highly motile also raises the averaged diffusion rate $\bar D_0$, and invasion speeds up if the product $\bar D_0\bar r_0$ increases. When death in phase $k$ is negligible, this occurs if
\begin{equation}
D_k>\left(2+\frac{\Delta t}{\tau}\right)\bar D_0.
\end{equation}
For a drug that doubles the cell-cycle duration, cells in the lengthened phase must be more than three times as motile as on average. This condition is difficult to satisfy (\cref{fig:lengthening-map}) because it requires lengthening a phase that takes a small fraction of the cycle and in which cells are highly motile. In the example of \cref{fig:lengthening}, $D_{\rm G1}<2\bar D_0$, so lengthening G1 never speeds up invasion.

In summary, the CCTD of treated cells alone predicts the same change in invasion speed whichever phase is lengthened. The error of this prediction grows with $\Delta t$ and with the difference between $D_k$ and $\bar D_0$. Closing this gap requires surprisingly little. If treatment lengthens a known phase without changing the phase-specific diffusion and death rates, one phase-resolved experiment on untreated cells is enough to predict the change in invasion speed for any treatment whose CCTD is known, without tracking treated cells.

%--------------------------------------------------------------------------------
\section{Discussion} \label{sec:discussion}
%--------------------------------------------------------------------------------

We were interested in how cell-cycle progression, motility and death, through their dependence on cell-cycle phase, determine the invasion speed in collective cell migration. Starting from a stochastic individual-based model, we derived a structured population model and established how the invasion speed depends on these cell-level rules. We expressed that speed in terms of quantities measured on single cells, particularly the CCTD, and identified which measurements are sufficient to predict it.

Our mathematical results extend existing theory for the invasion speed of cell populations. \Cref{thm:speed} gives an explicit formula for the invasion speed when the diffusion and death rates vary through the cell cycle, whereas previous results assumed that motility does not depend on phase \citep{gavagnin2019,abo2025survival}. For stochastic cell-cycle progression, we used the Feynman--Kac formula to express the second-order phase problem into a dispersion relation over the phase paths of single cells (\Cref{prop:stoch-speed}). The invasion speed then depends on the joint distribution of the cell-cycle time, how much a cell diffuses over its cycle and its survival. Both results provide exact formulae for the asymptotic invasion speed within a general framework of structured equations with renewal conditions, though well suited to modelling cell-cycle progression. At equal mean cell-cycle time and equal averaged rates, variability in cell-cycle times can only increase the invasion speed, for any level of phase noise, and the increase is of second order in the CV (\Cref{prop:cv}). When the rates do not depend on phase, the dispersion relation reduces to the Laplace transform of the CCTD, and we recover the result of \citet{gavagnin2019}.

Biologically, our results imply that the invasion speed cannot tell apart populations that move or die in different phases. Two populations invade at the same speed whenever their averaged diffusion and net growth rates, via the product $\bar D_0\bar r_0$, are the same. For example, C8161 and 1205Lu melanoma cells differ in diffusion rate and in cell-cycle duration, yet have nearly the same predicted invasion speed. The phase structure is not lost, however. It is present in the phase profile of the leading edge, which rises over phases in which cells are more motile than average (\cref{eq:edge-ratio}) and decreases otherwise. This separates two questions that are often asked together: how fast a population invades, and how motility is distributed over the cycle. The invasion speed can only answer the first question, whereas we have shown that the phase profile of the leading edge carries information on the second. Measuring this profile requires data that resolves cell-cycle phases. FUCCI imaging provides information on the spatial distribution of cell-cycle phases, but only at the resolution of the reporter. The typical two-colour FUCCI reporter resolves G1 and S/G2/M \citep{sakauesawano2008}; the four-colour FUCCI4 reporter resolves G1, S, G2 and M \citep{bajar2016fucci4}. 

The difficulty of distinguishing phase-specific motility from invasion speed alone is consistent with identifiability studies using FUCCI scratch assay data. \citet{simpson2020} fitted a reaction--diffusion model to FUCCI scratch assay data for 1205Lu melanoma cells and found that the diffusion rates of G1 and S/G2/M cells could not be separately estimated. Later, \citet{browning2023} showed that these rates are structurally identifiable from spatio-temporal measurements of the density of cells in each phase. So, the challenge is one of practical identifiability. \Cref{thm:speed} offers one interpretation whereby the invasion speed constrains the cycle-averaged diffusion rate, but not the diffusion rates of individual phases. Moreover, for 1205Lu cells, the diffusion rates estimated from single-cell trajectories are nearly equal in G1 and S/G2/M (\cref{tab:melanoma}), making them particularly difficult to distinguish experimentally. However, we note that scratch assays are not necessarily pulled fronts, and density profiles contain more information than invasion speed alone.

The invasion speed is also robust to variability in cell-cycle times. At measured CVs, variability changes it by a few percent. This is a strong prediction of the model, and it can be tested by comparing populations whose CCTDs have the same mean but different spreads, with the same averaged diffusion and death rates. For experimentalists, the take-away is that invasion speed depends mainly on the mean duration of each phase, and only weakly on the shapes of the phase-time distributions. Errors in the measured shapes, for example from censoring or small samples, do not matter as much as errors in the measured mean duration. For modellers, the conclusion is that deterministic cell-cycle progression with the correct averaged rates $\bar D_0$ and $\bar r_0$ is enough to predict the speed of a pulled front.

For treatments that lengthen the cell cycle, the CCTD of treated cells gives the increase in cell-cycle duration but not the phase that is lengthened. In our example, the CCTD alone predicts a 29\% reduction in invasion speed when the cell-cycle duration doubles, whereas lengthening G1 or S/G2/M reduces it by 19\% or 41\% (\cref{fig:lengthening}). How much this matters depends on how strongly motility depends on phase, and the evidence is mixed. Some cell lines move less in G2 than in G1/S \citep{walmod2004}, and invading cancer cells are mostly in G1 \citep{yano2014}. In contrast, cancer cells in S/G2/M can be more motile and invasive than cells in G1 \citep{kagawa2013cell}. In the three melanoma lines considered here, the diffusion rates in G1 and S/G2/M are similar \citep{vittadello2020}, so the CCTD alone would predict the change in invasion speed well. More generally, if treatment lengthens a known phase without changing the phase-specific diffusion and death rates, measurements on untreated cells are sufficient to predict its effect on invasion speed without tracking treated cells.

Together, these results identify what must be measured to predict the speed of a pulled invasion front. At first order, the mean duration of each phase and the diffusion and death rates of cells in each phase are enough. The shapes of the phase-time distributions contribute only at second order.

Our analysis has some limitations. Our results hold for pulled fronts, under the assumption that crowding slows cell-cycle progression and motility and increases death (\cref{eq:density-assumptions}). If cells instead proliferate faster at intermediate densities than at low density, fronts can become pushed, and the invasion speed then depends on the dynamics behind the leading edge \citep{vansaarloos2003,birzu2018}. We also assumed that each cell progresses through the cell cycle independently of its mother and its sister. Some data on cell-cycle times indicate correlations between related cells \citep{sandler2015,kuchen2020}, and we did not study how such correlations change the invasion speed. Importantly, the population-level model describes many cells, but in small populations, fluctuations at the leading edge change the speed of pulled fronts \citep{birzu2018}. It is therefore important to interpret our results in the context of a large ensemble of migrating cells. Within these limits, predicting the invasion speed of a cell population requires the mean time that cells spend in each phase, and how fast they move and die in each phase.

{
% \bmhead{Acknowledgements}

\section*{Statements and Declarations}
\begin{itemize}
\item \textbf{Funding} R.E.B. acknowledges support from the Simons Foundation (MP-SIP-00001828) and S.M.C.A. from a Royal Society Career Development Fellowship. 
\item \textbf{Competing interests} The authors declare no competing interests.
\item \textbf{Data availability} No new experimental data were generated. The data analysed are available from \citet{vittadello2020} and \citet{jones_2018}.
\item \textbf{Code availability} The code used to reproduce the figures and numerical simulations is available at \url{https://github.com/Stephanie-Abo/cell-cycle-pulled-invasion-speed}. Claude (Anthropic) was used to assist with figure-formatting code. All scientific content and analyses were developed and verified by the authors, who take full responsibility for the work.
\item \textbf{Author contributions} S.M.C.A.: Conceptualisation, Methodology, Software, Formal analysis, Investigation, Validation, Visualisation, Funding acquisition, Writing -- original draft, Writing -- review \& editing. R.E.B.: Conceptualisation, Methodology, Funding acquisition, Writing -- review \& editing.
\end{itemize}
}
%

%--------------------------------------------------------------------------------
\appendix
%--------------------------------------------------------------------------------
\section{Formal derivation of the population-level model}\label{app:meanfield}

We derive the population model \crefrange{eq:macro-pde}{eq:initial-condition} formally via a mean-field limit, following \citet{carrillo2010}, and treat division and death events as in \citet{tran2008}. 
% We first let $K\to\infty$ at fixed $\varepsilon$, and then let $\varepsilon\to0$.

We define the empirical measure
\[
    p^K(a,x,t)=\frac{1}{K}\sum_{i=1}^{M(t)}\delta(a-A_i(t))\,\delta(x-X_i(t)),
\]
where $\delta$ is the Dirac delta function. This is a finite positive measure whose total mass, $\int_\R\int_0^1p^K\D a\D x=M(t)/K$, fluctuates in time due to division and death. 

Fix $T>0$. We assume that $M(t)/K$ is uniformly bounded on $[0,T]$, and that $p^K(\cdot,\cdot,0)$ converges weakly to $p_0$.

Let us consider a test function $\phi\in C^2([0,1]\times\R)$ with compact support in $x$, with
\[
    \langle p^K,\phi\rangle=\frac{1}{K}\sum_{i=1}^{M(t)}\phi(A_i(t),X_i(t)).
\]
Between division and death events, cell $i$ follows \crefrange{eq:micro-sde-x}{eq:micro-sde-a}, with $v=v(A_i,N_\varepsilon)$, $D=D(A_i,N_\varepsilon)$ and $\sigma=\sigma(A_i)$ evaluated at its phase and at the local density $N_\varepsilon=N_\varepsilon(X_i(t),t)$,
\[
    \D A_i=v\D t+\sigma\D W_i^a,\qquad
    \D X_i=\partial_x  D\D t+\sqrt{2D}\D W_i^x,\qquad \Sigma=\frac12\sigma^2,
\]
where $\sqrt{2D}$ is defined since $D\geq0$ \assref{A4}. We apply It\^o's lemma, with $W_i^a$ and $W_i^x$ independent,
\[
    \D\phi(A_i,X_i)=\partial_a \phi\D A_i+\partial_x  \phi\D X_i+\frac12\sigma^2\partial_{aa}\phi\D t+\frac12(2D)\partial_{xx}\phi\D t.
\]
We substitute the cell dynamics and obtain
\[
    \D\phi(A_i,X_i)=\left[v\,\partial_a \phi+\Sigma\,\partial_{aa}\phi+\partial_x (D\,\partial_x \phi)\right]\D t+\D\xi_i,
\]
where we used $\partial_x D\,\partial_x \phi+D\,\partial_{xx}\phi=\partial_x (D\,\partial_x \phi)$, and where
\[
    \D\xi_i=\sigma\,\partial_a \phi\D W_i^a+\sqrt{2D}\,\partial_x \phi\D W_i^x.
\]
The death of cell $i$ removes its contribution to $\langle p^K,\phi\rangle$ at rate $\mu=\mu(A_i,N_\varepsilon)\geq0$ \assref{A5},
\[
    \D\langle p^K,\phi\rangle_{\mathrm{death},i}=-\frac{1}{K}\mu\,\phi(A_i,X_i)\D t-\frac{1}{K}\phi(A_i,X_i)\D\zeta_i,
\]
where the martingale $\zeta_i$, with zero mean, accounts for the randomness of the death time. A cell reaching $A_i=1$ divides and is replaced by two daughter cells at phase $0$,
\[
    \D\langle p^K,\phi\rangle_{\mathrm{division},i}=\frac{1}{K}\left[2\phi(0,X_i)-\phi(1,X_i)\right].
\]
Summing over the cells,
\[
    \D\langle p^K,\phi\rangle=\frac{1}{K}\sum_{i=1}^{M(t)}\left[\D\phi(A_i,X_i)-\mu\,\phi(A_i,X_i)\D t-\phi(A_i,X_i)\D\zeta_i\right]+\frac{1}{K}\sum_{i:\,A_i=1}\left[2\phi(0,X_i)-\phi(1,X_i)\right],
\]
where the last sum accounts for the cells that divide at time $t$. We substitute the It\^o expression and obtain
\[
\begin{aligned}
    \D\langle p^K,\phi\rangle={}&\frac{1}{K}\sum_{i=1}^{M(t)}\left[v\,\partial_a \phi+\Sigma\,\partial_{aa}\phi+\partial_x (D\,\partial_x \phi)-\mu\phi\right]\D t\\
    &+\frac{1}{K}\sum_{i=1}^{M(t)}\left[\D\xi_i-\phi\,\D\zeta_i\right]+\frac{1}{K}\sum_{i:\,A_i=1}\left[2\phi(0,X_i)-\phi(1,X_i)\right].
\end{aligned}
\]
The rates $v$, $D$ and $\mu$ depend on the local density, which we express in terms of the empirical measure,
\[
    N_\varepsilon(x,t)=\frac{1}{K}\sum_{j=1}^{M(t)}B_\varepsilon(x-X_j)=\frac{1}{K}\sum_{j=1}^{M(t)}\langle B_\varepsilon(x-y),\delta(y-X_j)\rangle_y=(B_\varepsilon*\rho^K)(x,t),
\]
where $\rho^K$ is the empirical cell density summed over all phases,
\[
    \rho^K(x,t)=\int_0^1p^K(a,x,t)\D a.
\]
The rates evaluated at cell $i$ are therefore
\[
    v(A_i,B_\varepsilon*\rho^K),\qquad D(A_i,B_\varepsilon*\rho^K),\qquad \mu(A_i,B_\varepsilon*\rho^K).
\]

Consider the empirical measure of division events,
\[
    \beta^K(x,t)=\frac{1}{K}\sum_{k}\delta(x-X_{i_k}(\tau_k))\,\delta(t-\tau_k),
\]
where cell $i_k$ divides at time $\tau_k$. The noise sum defines the martingale
\[
    \D\mathcal{M}^K=\frac{1}{K}\sum_{i=1}^{M(t)}\left[\D\xi_i-\phi\,\D\zeta_i\right],\qquad \mathcal{M}^K(0)=0.
\]

We write the drift sum in $\D\langle p^K,\phi\rangle$ as an integral with respect to the empirical measure,
\[
    \frac{1}{K}\sum_{i=1}^{M(t)}\left[v\,\partial_a \phi+\Sigma\,\partial_{aa}\phi+\partial_x (D\,\partial_x \phi)-\mu\phi\right]
    =\int_\R\int_0^1p^K\left[v\,\partial_a \phi+\Sigma\,\partial_{aa}\phi+\partial_x (D\,\partial_x \phi)-\mu\phi\right]\D a\D x,
\]
with the rates evaluated at $(a,B_\varepsilon*\rho^K)$. The division sum is an integral with respect to the empirical measure of division events,
\[
    \frac{1}{K}\sum_{i:\,A_i=1}\left[2\phi(0,X_i)-\phi(1,X_i)\right]
    =\int_\R\beta^K\left[2\phi(0,x)-\phi(1,x)\right]\D x.
\]

Integrating over $[0,t]$, we obtain
\[
\begin{aligned}
    \langle p^K(t),\phi\rangle={}&\langle p^K(0),\phi\rangle+\int_0^t\int_\R\int_0^1p^K\left[v\,\partial_a \phi+\Sigma\,\partial_{aa}\phi+\partial_x (D\,\partial_x \phi)-\mu\phi\right]\D a\D x\D s\\
    &+\int_0^t\int_\R\beta^K\left[2\phi(0,x)-\phi(1,x)\right]\D x\D s+\mathcal{M}^K(t).
\end{aligned}
\]

We now let $K\to\infty$ at fixed $\varepsilon$, term by term. The first term converges to $\langle p_0,\phi\rangle$ by assumption.

For the drift term, we assume that $p^K$ converges weakly to a deterministic limit with a smooth density $p^\varepsilon(a,x,t)$, so that $\rho^K=\int_0^1p^K\D a$ converges weakly to $\rho^\varepsilon=\int_0^1p^\varepsilon\D a$. Since $B_\varepsilon$ is smooth for $\varepsilon>0$, the local density and its derivative, which enters through $\partial_x  D$, converge pointwise,
\[
    (B_\varepsilon*\rho^K)(x,t)=\int_\R B_\varepsilon(x-y)\,\rho^K(y,t)\D y\;\to\;(B_\varepsilon*\rho^\varepsilon)(x,t),\qquad
    \partial_x (B_\varepsilon*\rho^K)\;\to\;\partial_x (B_\varepsilon*\rho^\varepsilon).
\]
Since $v$, $D$ and $\mu$ are continuous in $N$, the rates converge. The drift term therefore converges to the same integral with $p^\varepsilon$ in place of $p^K$, and the rates evaluated at $(a,N_\varepsilon)$, where now $N_\varepsilon=B_\varepsilon*\rho^\varepsilon$.

For the division term, we assume that $\beta^K$ converges weakly to a rate density $b^\varepsilon(x,t)$, so that it converges to
\[
    \int_0^t\int_\R b^\varepsilon\left[2\phi(0,x)-\phi(1,x)\right]\D x\D s.
\]

The noise term is a sum of uncorrelated martingales. Its variance is $O(1/K)$, so it vanishes as $K\to\infty$.

Collecting the limits and differentiating in $t$, we obtain
\[
    \frac{\D}{\D t}\langle p^\varepsilon,\phi\rangle=\int_\R\int_0^1p^\varepsilon\left[v\,\partial_a \phi+\Sigma\,\partial_{aa}\phi+\partial_x (D\,\partial_x \phi)-\mu\phi\right]\D a\D x+\int_\R b^\varepsilon\left[2\phi(0,x)-\phi(1,x)\right]\D x,
\]
with the rates evaluated at $(a,N_\varepsilon)$.

Since $p^\varepsilon$ is smooth, we integrate by parts term by term. In $x$, we obtain
\[
    \int_\R p^\varepsilon\,\partial_x (D\,\partial_x \phi)\D x
    =\left[p^\varepsilon\,D\,\partial_x \phi\right]_{-\infty}^{\infty}-\int_\R D\,\partial_x p^\varepsilon\,\partial_x \phi\D x
    =-\left[D\,\partial_x p^\varepsilon\,\phi\right]_{-\infty}^{\infty}+\int_\R\partial_x (D\,\partial_x p^\varepsilon)\,\phi\D x,
\]
where the boundary terms vanish since $\phi$ has compact support in $x$.

In $a$, we treat the transport and diffusion terms separately. The transport term gives
\[
    \int_0^1p^\varepsilon\,v\,\partial_a \phi\D a
    =\left[v\,p^\varepsilon\,\phi\right]_0^1-\int_0^1\partial_a (v\,p^\varepsilon)\,\phi\D a,
\]
and the diffusion term, integrated by parts twice, gives
\[
    \int_0^1p^\varepsilon\,\Sigma\,\partial_{aa}\phi\D a
    =\left[\Sigma\,p^\varepsilon\,\partial_a \phi\right]_0^1-\int_0^1\partial_a (\Sigma\,p^\varepsilon)\,\partial_a \phi\D a
    =-\left[\partial_a (\Sigma\,p^\varepsilon)\,\phi\right]_0^1+\int_0^1\partial_{aa}(\Sigma\,p^\varepsilon)\,\phi\D a,
\]
where the first bracket vanishes since $\Sigma(0)=\Sigma(1)=0$ \assref{A2}. Introducing the phase flux $J^\varepsilon=v\,p^\varepsilon-\partial_a (\Sigma\,p^\varepsilon)$, the two combine as
\[
\begin{aligned}
    \int_0^1p^\varepsilon\left(v\,\partial_a \phi+\Sigma\,\partial_{aa}\phi\right)\D a
    &=\left[J^\varepsilon\phi\right]_0^1-\int_0^1\partial_a J^\varepsilon\,\phi\D a\\
    &=J^\varepsilon(1,x,t)\,\phi(1,x)-J^\varepsilon(0,x,t)\,\phi(0,x)-\int_0^1\partial_a J^\varepsilon\,\phi\D a.
\end{aligned}
\]
The phase endpoints are not restricted by the support of $\phi$, so these boundary terms are retained. Collecting, we obtain
\[
\begin{aligned}
    \frac{\D}{\D t}\langle p^\varepsilon,\phi\rangle={}&\int_\R\int_0^1\left[\partial_x (D\,\partial_x p^\varepsilon)-\partial_a J^\varepsilon-\mu\,p^\varepsilon\right]\phi\D a\D x\\
    &+\int_\R\left[J^\varepsilon(1,x,t)\,\phi(1,x)-J^\varepsilon(0,x,t)\,\phi(0,x)+b^\varepsilon(x,t)\left(2\phi(0,x)-\phi(1,x)\right)\right]\D x.
\end{aligned}
\]
The number of cells reaching the dividing phase per unit time at position $x$ is the phase flux at $a=1$, so $b^\varepsilon(x,t)=J^\varepsilon(1,x,t)$. The boundary terms then combine as
\[
\begin{aligned}
    &J^\varepsilon(1,x,t)\,\phi(1,x)-J^\varepsilon(0,x,t)\,\phi(0,x)+2J^\varepsilon(1,x,t)\,\phi(0,x)-J^\varepsilon(1,x,t)\,\phi(1,x)\\
    &\qquad=\left[2J^\varepsilon(1,x,t)-J^\varepsilon(0,x,t)\right]\phi(0,x).
\end{aligned}
\]
Since $\phi$ is arbitrary, we obtain
\begin{equation}\label{eq:app-pde-eps}
    \partial_tp^\varepsilon+\partial_a J^\varepsilon=\partial_x \left(D(a,N_\varepsilon)\,\partial_x p^\varepsilon\right)-\mu(a,N_\varepsilon)\,p^\varepsilon,\qquad N_\varepsilon=B_\varepsilon*\int_0^1p^\varepsilon\D a,
\end{equation}
and the boundary term gives the renewal condition $J^\varepsilon(0,x,t)=2J^\varepsilon(1,x,t)$, with $p^\varepsilon(a,x,0)=p_0(a,x)$.

Finally, we let $\varepsilon\to0$. We assume that the solution $p^\varepsilon$ of \cref{eq:app-pde-eps} converges locally uniformly to $p$, together with its derivatives, so that $\rho^\varepsilon=\int_0^1p^\varepsilon\D a\to N$, with $N$ given by \cref{eq:local-density-limit}. Since $B_\varepsilon\geq0$ has unit integral and support in $[-\varepsilon,\varepsilon]$, $N_\varepsilon=B_\varepsilon*\rho^\varepsilon\to N$ and $\partial_x  N_\varepsilon\to\partial_x  N$. This recovers the model \crefrange{eq:macro-pde}{eq:initial-condition}, independently of the shape of $B$.

This derivation is formal. For a rigorous derivation of the limit $K\to\infty$ in a related model with spatial diffusion, birth and death, see \citet{champagnat2007}.

%--------------------------------------------------------------------------------
\section{Proof of \cref{thm:speed}}\label{app:speed-deterministic}
%--------------------------------------------------------------------------------

\begin{proof}
Substituting $p(a,z)=\varphi(a)e^{-\lambda z}$ into \cref{eq:edge-problem} gives \cref{eq:edge-ode}, with the renewal condition $v_0(0)\varphi(0)=2v_0(1)\varphi(1)$. \Cref{eq:edge-ode} is a first-order linear equation for $v_0\varphi$, with solution
\[
    v_0(a)\varphi(a)=v_0(0)\varphi(0)\exp\left(\int_0^a\frac{D_0(s)\lambda^2-c\lambda-\mu_0(s)}{v_0(s)}\D s\right).
\]
We note that $\varphi$ is positive whenever $\varphi(0)>0$. Evaluating at $a=1$ and substituting the renewal condition, we obtain
\[
    1=2\exp\left(\int_0^1\frac{D_0(a)\lambda^2-c\lambda-\mu_0(a)}{v_0(a)}\D a\right).
\]
Taking logarithms and using \cref{eq:cycle-total}, we obtain the dispersion relation \cref{eq:F0},
\[
    F_0(c,\lambda)=\tot{D_0}\lambda^2-c\tau\lambda+\ln 2-\tot{\mu_0}=0.
\]
A positive solution of the form $\varphi(a)e^{-\lambda z}$ exists if and only if $F_0$ has a positive real root $\lambda$. Since $\tot{D_0}>0$ and $\ln 2>\tot{\mu_0}$, and with $c>0$, any real root is then positive. The roots are real when the discriminant is non-negative,
\[
    c^2\tau^2-4\tot{D_0}\left(\ln 2-\tot{\mu_0}\right)\geq0,
\]
that is, 
\[
    c\geq\cmin=2\sqrt{\tot{D_0}(\ln 2-\tot{\mu_0})}/\tau=2\sqrt{\bar D_0\bar r_0}. 
\]
At $c=\cmin$, the discriminant vanishes and the two roots coalesce at $\lambda_c=\sqrt{(\ln 2-\tot{\mu_0})/\tot{D_0}}=\sqrt{\bar r_0/\bar D_0}$.
\end{proof}

%--------------------------------------------------------------------------------
\section{Proof of \cref{prop:stoch-speed}}\label{app:speed-stochastic}
%--------------------------------------------------------------------------------

\begin{proof}
Substituting $p(a,z)=\varphi(a)e^{-\lambda z}$ into \cref{eq:edge-problem-sigma} gives the second-order phase problem
\[
    \partial_a  J_\varphi=\left(D_0(a)\lambda^2-c\lambda-\mu_0(a)\right)\varphi,\qquad J_\varphi:=v_0(a)\varphi-\partial_a \left(\Sigma(a)\varphi\right),\qquad J_\varphi(0)=2J_\varphi(1).
\]
Fix $\lambda$ and $c$, and let $\psi(a)$ solve the adjoint of the phase problem
\begin{equation}\label{eq:adjoint}
    \Sigma(a)\,\partial_{aa}\psi+v_0(a)\,\partial_a \psi+\left(D_0(a)\lambda^2-c\lambda-\mu_0(a)\right)\psi=0,\qquad \psi(1)=1,
\end{equation}
with $\psi$ bounded as $a\to0$ (\assref{A2} and \assref{A3}). We multiply the phase problem by $\psi$ and integrate over one cycle,
\[
    \int_0^1\partial_aJ_\varphi\,\psi\D a=\int_0^1\left(D_0\lambda^2-c\lambda-\mu_0\right)\varphi\,\psi\D a.
\]
We integrate the left-hand side by parts and use the definition of $J_\varphi$,
\[
    \int_0^1\partial_aJ_\varphi\,\psi\D a
    =\Big[J_\varphi\psi\Big]_0^1-\int_0^1v_0\,\varphi\,\partial_a\psi\D a+\int_0^1\partial_a(\Sigma\varphi)\,\partial_a\psi\D a.
\]
Then we integrate the last term by parts again,
\[
    \int_0^1\partial_a(\Sigma\varphi)\,\partial_a\psi\D a=\Big[\Sigma\varphi\,\partial_a\psi\Big]_0^1-\int_0^1\Sigma\varphi\,\partial_{aa}\psi\D a.
\]
Collecting the terms, we obtain
\[
    0=\Big[J_\varphi\psi+\Sigma\varphi\,\partial_a \psi\Big]_{a=0}^{a=1}-\int_0^1\varphi\left(\Sigma\,\partial_{aa}\psi+v_0\,\partial_a \psi+\left(D_0\lambda^2-c\lambda-\mu_0\right)\psi\right)\D a.
\]
The integral vanishes because $\psi$ solves \cref{eq:adjoint}, and the second boundary term vanishes because $\Sigma(0)=\Sigma(1)=0$~\assref{A2}. The renewal condition then gives $J_\varphi(1)\left(\psi(1)-2\psi(0)\right)=0$. For a positive solution, the flux of dividing cells $J_\varphi(1)=\left(v_0(1)-\partial_a \Sigma(1)\right)\varphi(1)$ is positive, since $\Sigma\geq0$ and $\Sigma(1)=0$ imply $\partial_a \Sigma(1)\leq0$~\assref{A2}. Since $\psi(1)=1$,
\begin{equation}\label{eq:cycle-weight-renewal}
    2\psi(0)=1.
\end{equation}
The bounded solution of \cref{eq:adjoint} has the Feynman--Kac representation, since $T$ is finite almost surely~\assref{A1}
\begin{equation}\label{eq:cycle-weight}
    \psi(0)=\E\left[\exp\left(\int_0^T\left(D_0(A(t))\lambda^2-c\lambda-\mu_0(A(t))\right)\D t\right)\right],
\end{equation}
where $A(t)$ is the cell-cycle phase of a newborn cell at low density, given by \cref{eq:micro-sde-a} with cycling speed $v_0(a)$, $A(0)=0$, and $T$ its first-passage time to $A=1$. Substituting \cref{eq:cycle-weight} into \cref{eq:cycle-weight-renewal} gives \cref{eq:disp-stoch}. 

We note that $\psi(0)$ in \cref{eq:cycle-weight} is strictly decreasing in $c$ since $\lambda T>0$, so \cref{eq:cycle-weight-renewal} has at most one solution $c(\lambda)$ for each $\lambda>0$. At $c=0$, $\psi(0)\geq\E\left[e^{-\int_0^T\mu_0(A(t))\D t}\right]$ as $\lambda^2\int_0^T D_0(A(t))\D t\geq0$. Since $2\E[e^{-\int_0^T\mu_0(A(t))\D t}]>1$ by assumption, $2\psi(0)>1$ at $c=0$, and the solution to \cref{eq:cycle-weight-renewal} is positive. The invasion speed is the minimum of $c(\lambda)$ over $\lambda>0$.
\end{proof}

\begin{remark}
    When $D_0(a)\equiv D$ and $\mu_0(a)\equiv\mu$ do not depend on cell-cycle phase, $\psi(0)$ is the Laplace transform of the CCTD,
\[
    \mathcal{L}\{f_T\}(s)=\E\left[e^{-sT}\right],\qquad s=c\lambda-D\lambda^2+\mu,
\]
and the dispersion relation reduces to $2\mathcal{L}\{f_T\}(s)=1$.
\end{remark}
%

% --------------------------------------------------------------------------------
\section{Proof of \cref{prop:cv}}\label{app:cv}
%--------------------------------------------------------------------------------

\begin{proof}
Assume that $\Sigma$ is small. As in \cref{prop:cv}, we compare populations at equal means, so $F_0$ is evaluated with the means over one cycle, and $\tau=\E[T]$. Let
\[
    X=\lambda^2\int_0^T D_0(A(t))\D t-c\lambda T-\int_0^T\mu_0(A(t))\D t,
\]
where $A(t)$ is the phase at time $t$, so \cref{eq:disp-stoch} reads $2\E[e^X]=1$. With $g_0(a)=(D_0(a)\lambda^2-c\lambda-\mu_0(a))/v_0(a)$, $X=\int_0^T v_0(A(t))\,g_0(A(t))\D t$.

By \cref{eq:micro-sde-a}, $\D A(t)=v_0(A(t))\D t+\sigma(A(t))\D W(t)$. We apply It\^o's lemma to $\int_0^{A(t)}g_0(s)\D s$, with $\Sigma=\sigma^2/2$, and obtain
\[
    \D\int_0^{A(t)}g_0(s)\D s=v_0(A(t))\,g_0(A(t))\D t+\Sigma(A(t))\,\partial_a g_0(A(t))\D t+g_0(A(t))\,\sigma(A(t))\D W(t).
\]
We integrate from $t=0$ to $T$. Since $A(0)=0$ and $A(T)=1$, the left-hand side gives $\int_0^1g_0(a)\D a$, and the first term on the right gives $X$. Hence
\[
    X=\int_0^1 g_0(a) \D a-\int_0^T\Sigma(A(t))\,\partial_a g_0(A(t))\D t-\int_0^T g_0(A(t))\,\sigma(A(t))\D W(t).
\]

By the It\^o isometry and to leading order in $\Sigma$,
\[
    \mathrm{Var}(X)=\E\int_0^T2\Sigma(A(t))\,g_0(A(t))^2\D t+O(\Sigma^2)=\int_0^1\frac{2\Sigma}{v_0}\,g_0^2\D a+O(\Sigma^2)=2F_1+O(\Sigma^2),
\]
with $F_1$ as in \cref{eq:cv-general}.

We write $X= \E[X] + (X-\E[X])$ and expand about the mean. Since $\E[X-\E[X]]=0$ and the higher moments of $X-\E[X]$ are $O(\Sigma^2)$,
\[
    \E[e^X]=e^{\E[X]}\,\E\left[e^{X-\E[X]}\right]=e^{\E[X]}\left(1+\tfrac{1}{2}\mathrm{Var}(X)+O(\Sigma^2)\right).
\]
Taking the logarithm of $2\E[e^X]=1$,
\[
    \ln2+\E[X]+\ln\left(1+\tfrac{1}{2}\mathrm{Var}(X)+O(\Sigma^2)\right)=0.
\]
Since $\mathrm{Var}(X)=O(\Sigma)$, we obtain
\[
\ln2+\E[X]+\tfrac12\mathrm{Var}(X)+O(\Sigma^2)=0,
\]
where the first two terms are $F_0$,
\[
    \ln2+\E[X]= \ln2 +\E\left[\int_0^T D_0(A(t))\D t\right]\lambda^2-c\lambda\,\E[T]-\E\left[\int_0^T\mu_0(A(t))\D t\right].
\]
Hence
\[
    F_0+F_1+O(\Sigma^2)=0.
\]

Phase noise therefore enters only through $F_1$.

Since $\partial_\lambda F_0=0$ at the double root $(\cmin,\lambda_c)$ of $F_0$, we linearise in $c$ at fixed $\lambda_c$, with $\partial_c F_0=-\lambda\,\E[T]$,
\[
    -\lambda_c\,\E[T]\left(c_\ast-\cmin\right)+F_1(\cmin,\lambda_c)=O(\Sigma^2),
\]
which gives \cref{eq:cv-general}.

\emph{Constant rates.} For $D_0\equiv D$ and $\mu_0\equiv\mu$, $X=(D\lambda^2-c\lambda-\mu)\,T$ and $F_0=(D\lambda^2-c\lambda-\mu)\,\E[T]+\ln2$. At the double root, $F_0=0$ gives $D\lambda_c^2-\cmin\lambda_c-\mu=-\ln2/\E[T]$, so $X=-\ln2\,T/\E[T]$. Since $F_1=\tfrac12\mathrm{Var}(X)+O(\Sigma^2)$,
\[
    F_1=\tfrac12(\ln2)^2\,\frac{\mathrm{Var}(T)}{\E[T]^2}+O(\Sigma^2)=\tfrac12(\ln2)^2\,\mathrm{CV}^2+O(\Sigma^2),
\]
by the definition of the CV. Substituting into \cref{eq:cv-general} gives \cref{eq:cv-constant}.

\emph{Order of the CV.} The main text compares $F_1$ with $\mathrm{CV}^2=\mathrm{Var}(T)/\E[T]^2$. The cycle time $T=\int_0^T v_0(A(t))\,\tfrac{1}{v_0(A(t))}\D t$ has the form of $X$, with $1/v_0$ in place of $g_0$, so the same computation of $\mathrm{Var}(X)$ above gives $\mathrm{Var}(T)=\int_0^1 2\Sigma/v_0^3\D a+O(\Sigma^2)$. Hence $\mathrm{CV}^2=O(\Sigma)$, which is the same order as $F_1$.
\end{proof}

\begin{remark}[Lower bound for any $\Sigma$]\label{rem:lower-bound}
For any phase noise, $c_\ast\geq\cmin$, provided that $\E\big[\int_0^T\mu_0(A(t))\D t\big]<\ln2$. With $X$ as in the proof above and evaluated at $c=c(\lambda)$, \cref{eq:disp-stoch} and Jensen's inequality give
\[
    1=2\E[e^X]\geq2e^{\E[X]},\qquad\text{so}\qquad F_0(c(\lambda),\lambda)=\ln2+\E[X]\leq0.
\]
By the proof of \Cref{thm:speed} (Appendix~\ref{app:speed-deterministic}), $F_0(c,\lambda)>0$ for all $\lambda>0$ whenever $c<\cmin$. Hence $c(\lambda)\geq\cmin$ for every $\lambda>0$, and $c_\ast=\min_\lambda c(\lambda)\geq\cmin$.
\end{remark}

%--------------------------------------------------------------------------------
\section{Experimental data analysis}\label{app:data-analysis}
%--------------------------------------------------------------------------------

We use the single-cell data of \citet{vittadello2020} for three FUCCI-labelled melanoma cell lines, C8161, WM983C and 1205Lu. Cells were seeded at $10^4$ cells\,cm$^{-2}$ in 2-D proliferation assays and imaged every 15\,min for 48\,h. It is a subconfluent culture, so cells move freely at low density. For each cell line, the data contain 50 trajectories of cells in G1 and 50 trajectories of cells in S/G2/M, sampled at random from more than $10^3$ trajectories. Each trajectory follows one cell for the complete duration of one phase.

For each phase, we estimate the diffusion rate from the net displacements of cells using
\begin{equation}
D_{\rm{phase}}:=\frac{\sum_j \left[\left(x_j(T_j)-x_j(0)\right)^2+\left(y_j(T_j)-y_j(0)\right)^2\right]}{4\sum_j T_j},
\end{equation}
where $T_j$ is the time that cell $j$ spends in the phase, and the factor 4 accounts for movement in two dimensions. We take the mean durations of G1 and S/G2/M, $\tau_{\rm G1}$ and $\tau_{\rm S/G2/M}$, from the same trajectories. A trajectory covers a complete phase only if that phase starts and ends within the 48\,h of imaging, which may bias the measured durations towards short phases. We therefore rescale them and compute the time-averaged diffusion rate as
\begin{equation}
\bar D_0=\frac{D_{\rm G1}\tau_{\rm G1}+D_{\rm S/G2/M}\tau_{\rm S/G2/M}}{\tau_{\rm G1}+\tau_{\rm S/G2/M}}.
\end{equation}
The mean-square displacement along one axis over one cycle is $2\tot{D_0}=2\bar D_0\tau$. 

Death rates were not measured, so we take $\bar r_0=\ln2/\tau$. The invasion speed follows from \cref{eq:cmin}. Since each trajectory covers one phase only, the data do not give the cell-cycle duration $\tau$, and we use the duration reported in Table S2 of \citet{vittadello2020}. The mean phase durations we computed do not sum to this $\tau$, and neither do those reported in Table S2. For C8161, $\ln2/\tau=0.033$\,h$^{-1}$ agrees with the net growth rate of 0.036\,h$^{-1}$ obtained from the cell counts reported by \citet{vittadello2020}, which motivated our modelling choice. \Cref{tab:melanoma} summarises the results. In \cref{fig:averages}e, error bars show $\tau$ $\pm$ one standard deviation and the 95\% bootstrap intervals over the trajectories for the mean-square displacement.

\begin{table}[ht]
\centering
\begin{tabular}{lcccccc}
\toprule
 & $\tau$ (h) & $D_{\rm G1}$ & $D_{\rm S/G2/M}$ & $\bar D_0$ & $\bar r_0$ (h$^{-1}$) & $\cmin$ ($\mu$m\,h$^{-1}$) \\
\midrule
C8161  & 21 & 51 & 39 & 46 & 0.033 & 2.45 \\
WM983C & 23 & 30 & 25 & 27 & 0.030 & 1.80 \\
1205Lu & 37 & 76 & 78 & 77 & 0.019 & 2.40 \\
\botrule
\end{tabular}
\caption{Cell-cycle durations, diffusion rates ($\mu$m$^2$\,h$^{-1}$) and predicted invasion speeds of three melanoma cell lines, from the data of \citet{vittadello2020}.}
\label{tab:melanoma}
\end{table}

To include cell-cycle-time variability, we use the CV of the cell-cycle time reported in Table S2 of \citet{vittadello2020}. We use the averaged diffusion rate $\bar D_0$ with no death, and \cref{eq:cv-constant} gives $c_\ast/\cmin=1+(\ln2/4)\,\mathrm{CV}^2$. \Cref{tab:melanoma-cv} compares $c_\ast$ with $\cmin$.

\begin{table}[ht]
\centering

\begin{tabular}{lcccc}
\toprule
 & CV & $\cmin$ ($\mu$m\,h$^{-1}$) & $c_\ast$ ($\mu$m\,h$^{-1}$) & increase \\
\midrule
C8161  & 0.38 & 2.45 & 2.51 & 2.5\% \\
WM983C & 0.43 & 1.80 & 1.86 & 3.3\% \\
1205Lu & 0.43 & 2.40 & 2.48 & 3.2\% \\
\botrule
\end{tabular}
\caption{Effect of cell-cycle-time variability on the predicted invasion speeds of the three melanoma cell lines.}
\label{tab:melanoma-cv}
\end{table}

\section{Lengthening a cell-cycle phase}\label{app:lengthening}
\setcounter{figure}{0}
\gdef\thefigure{S\arabic{figure}}

\begin{figure}[ht]
\centering
\includegraphics[width=0.5\linewidth]{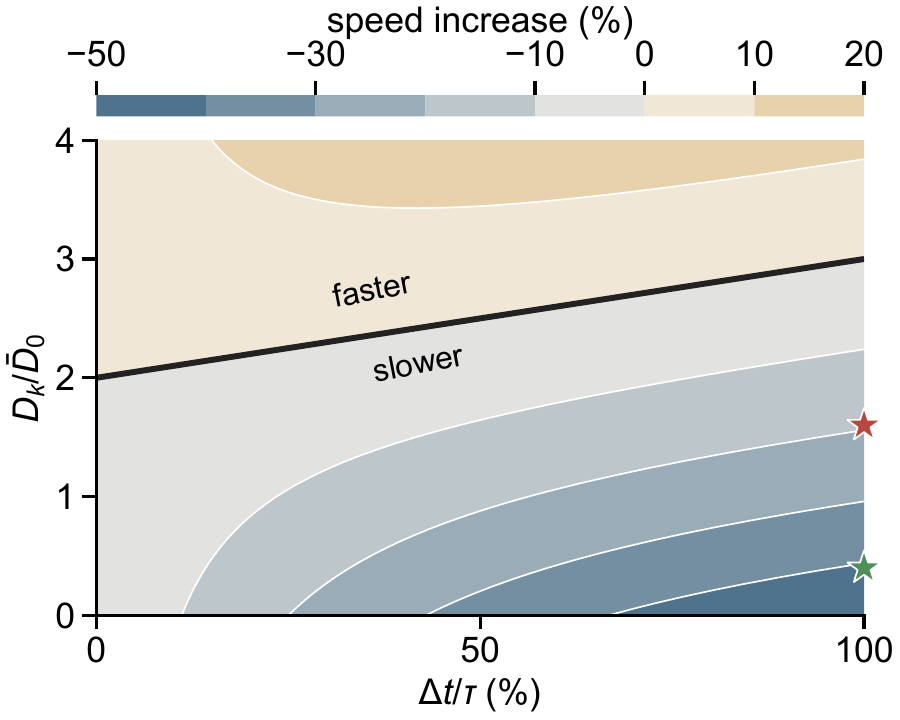}
\caption{Change in invasion speed when phase $k$ is lengthened by $\Delta t$, from \cref{eq:lengthening} with $\mu_k=0$. We plot the change in invasion speed as a function of the relative increase in cell-cycle duration, $\Delta t/\tau$, and of the ratio $D_k/\bar D_0$ of the diffusion rate in phase $k$ to the averaged diffusion rate before treatment. Invasion speeds up above the black line, $D_k=(2+\Delta t/\tau)\bar D_0$, and slows down below it. Each colour band spans 10\%. Stars mark the cases of \cref{fig:lengthening} in which G1 (red) or S/G2/M (green) is lengthened.}
\label{fig:lengthening-map}
\end{figure}

\Cref{fig:lengthening-map} shows the change in invasion speed given by \cref{eq:lengthening} when a phase $k$ in which death is negligible is lengthened by $\Delta t$.

\bibliography{references}

\end{document}